\documentstyle[epsf,12pt]{article}
\newcommand{\y }{\'{\i}}
\nonstopmode

\begin{document}

\begin{center}
\large {\bf KINETICS OF A MIXED ISING FERRIMAGNETIC SYSTEM}  
\end{center}
\vskip 16 truept
\centerline{\bf G. M. Buend\y a}   
\vskip 3 truept
\begin{center}
{Departamento de F\y sica, Universidad Sim\'on Bol\y var \\
 Apartado 89000, Caracas 1080, Venezuela}
\end{center}
\vskip 5 truept
\centerline{\bf E. Machado}   
\vskip 3 truept
\begin{center}
{Departamento de F\y sica, Universidad Sim\'on Bol\y var \\
 Apartado 89000, Caracas 1080, Venezuela \\
 and \\
 Departamento de F\y sica \\
 Universidad Central de Las Villas, Santa Clara 4300, Cuba}
\end{center}

\begin{abstract}
We present a study, within a mean-field approach, of the kinetics of a
classical mixed Ising ferrimagnetic model on a square lattice, in which the
two interpenetrating square sublattices have spins $\sigma = \pm1/2$
and $S = \pm 1,0$. The kinetics is described by a Glauber-type stochastic
dynamics in the presence of a time-dependent oscillating external field and
a crystal field interaction.
We can identify two types of solutions: a symmetric one, where the total
magnetization, $M$,
oscillates around zero, and an antisymmetric one where $M$ oscillates around
a finite value different from zero. There are regions of the phase space where
both solutions coexist. The dynamical transition from one regime to the other
can be of first or second order depending on the region in the phase diagram.
Depending on the value of the crystal field we
found up to two dynamical tricritical points where the transition changes from
continuous to discontinuous. Also, we perform a similar study on the
Blume-Capel ($S=\pm 1,0$) model and found strong differences between its
behavior and the one of the mixed model.
\end{abstract}

\bigskip
\leftline {\bf I. Introduction}

The time evolution of metaestable states of a mixed Ising ferrimagnetic
model are studied by establishing a Glauber-type dynamic~\cite{Glauber}
that drives
the system between two equivalent ordered phases. This approach provides
a simple way to introduce a nonequilibrium kinetics into models such as
the Ising and the Ising mixed systems, that do not have a deterministic
dynamics. Numerous studies indicate that stochastic dynamics, despite
it simplicity, can describe in a qualitative correct way nonequilibrium
phenomena and dynamical phase transitions found in real systems
~\cite{Acha}. Recently
a kinetic Ising model has been applied sucesfully to model magnetization
switching in nanoscale ferromagnets~\cite{Rikvold}.

The model to be studied is a mixed Ising ferrimagnetic system on a
square lattice in which the two interpenetrating sublattices have
spins one-half ($\pm1/2$) and spins one ($\pm1,0$). This system is
relevant for understanding bimetallic molecular ferrimagnets that are
currently being synthesized by several experimental groups in search
of stable, crystalline materials, with spontaneus magnetic moments at
room temperature~\cite{Kahn}. The magnetic properties of the mixed Ising models have
been studied by high-temperatures series expansions~\cite{Hunter}, 
renormalization-group~\cite{Bowers}, mean-field~\cite{Kaneyoshi}, Monte Carlo
simulations and numerical transfer matrix calculations~\cite{Buendia},
but always at equilibrium conditions. 

In this work we are going to study, within a mean field approach, the kinetics
of de model in presence of a time-dependent oscillating external field. We
found that there are three regimes: a paramagnetic one were the total
magnetization follows the field, a ferromagnetic one where the total
magnetization does not follows the field and oscillates around a value
different from zero, and a region where both solutions coexist. The transition
between regimes can be continuous or discontinuous, depending on the temperature
and the external field. Depending on the parameters of the Hamiltonian this
system can have up to two dynamical tricritical points.

\bigskip
\leftline {\bf II. The Model}

We consider a kinetic mixed Ising ferrimagnetic system in a square lattice
described by the Hamiltonian
\begin{equation}
{\cal H} = -J \sum_{\langle nn\rangle} S_i \sigma_j - D \sum_i S_i^2
- H( \sum_i S_i + \sum_j \sigma_j)
\end{equation}
where the $S_i$ take the values $\pm1$ or $0$ and are located in
alternating sites with spins $\sigma_j=\pm1/2$. Each spin $S$ has
only $\sigma$-spins as nearest-neighbors 
and vice-versa. The sum $\sum_{<nn>}$ is carried out over all
nearest-neighbors pairs. The sums $\sum_i$ and $\sum_j$ run over all the
sites of the $S$ and $\sigma$ sublattices, respectively. $J$ is the
exchange parameter, $D$ is the crystal field interaction,
and $H$ is a time-dependent external magnetic field given by 
\begin{equation}
H(t) = H_0 \cos(\omega t)\ ,
\label{field}
\end{equation}
all in energy units. We choose $J$ to be negative, so the coupling
between the nn spins is antiferromagnetic.

The system evolves according to a Glauber-type stochastic dynamics
~\cite{Glauber}, that is described a continuation,
at a rate of $1/\tau$ transitions per unit time. Leaving the
$S$-spins fixed, we define $P'( \sigma_1, \ldots , \sigma_N; t)$
as the probability that the system has the $\sigma$-spins
configuration, $\sigma_1, \ldots , \sigma_N$, at time $t$, also, by
leaving the $\sigma$-spins fixed, we define $P''(S_1, \ldots , S_N; t)$
as the probability that the system has the $S$-spins configuration,
$S_1, \ldots , S_N$, at time $t$. Then, we calculate $w_j(\sigma_j)$
and $W_i(S_i \rightarrow S'_i)$, the probabilities per unit time 
that the $j$th $\sigma$-spin changes from $\sigma_j$ to $- \sigma_j$
and the $i$th $S$-spin changes from $S_i$ to $S'_i$, respectively.

\bigskip
\leftline {\bf II.a. Calculation of $w_j(\sigma_j)$}

We write the time derivative of $P'( \sigma_1, \ldots , \sigma_N; t)$ as
\begin{eqnarray}
\frac{d}{dt}P'( \sigma_1, \ldots , \sigma_N; t)=
-[\sum_j w_j(\sigma_j)]P'( \sigma_1, \ldots , \sigma_j, \ldots , \sigma_N; t) 
\nonumber \\
+\sum_j w_j(-\sigma_j)
P'( \sigma_1, \ldots , -\sigma_j, \ldots , \sigma_N; t)\ .
\end{eqnarray}

If the system is in contact with a heat bath at temperature T,
each spin $\sigma$ can flip with probability per unit time given by the
Boltzmann factor~\cite{Binder} 
\begin{equation}
w_j(\sigma_j)=\frac{1}{\tau}
\frac{\exp(-\beta\Delta E'_j)}{1+\exp(-\beta\Delta E'_j)}\ \ \ ;\ \ \ 
\beta = 1/ k_BT
\end{equation}
where 
\begin{equation}
\Delta E'_j=2 \sigma_j(J \sum_{<i>} S_i + H)
\end{equation}
gives the change in the energy of the system when the $\sigma_j$ 
spin flips.

From the master equation associated to the stochastic process it is simple to
prove that the average $<\sigma_j(t)>$ satisfies the
following equation~\cite{Kubo}
\begin{equation}
\tau \frac{d}{dt} <\sigma_j>=-<\sigma_j>+
<\frac{1}{2} \tanh[\frac{1}{2} \beta (J \sum_{<i>} S_i + H)]>\ ,
\end{equation}
that, within a mean-field approach~\cite{Tome} and
for an external field defined by (\ref{field}), takes the form
\begin{equation}
\tau \frac{d}{dt} <\sigma>=-<\sigma>+ \frac{1}{2}
\tanh{\frac{1}{2} \beta [J Z <S>+H_0\cos(\omega t)]}\ ,
\label{m1}
\end{equation}
where $<\sigma>$ and $<S>$ are the $\sigma$-spin and $S$-spin
sublattice magnetizations respectively, and $Z=4$ is the coordination number
for this model.

\bigskip
\leftline {\bf II.b. Calculation of $W_i(S_i \rightarrow S'_i)$}

In a similar way, the time derivative of the
$P''(S_1, \ldots , S_N; t)$ can be written as
\begin{eqnarray}
\frac{d}{dt}P''(S_1, \ldots , S_N; t)=
-\sum_i [\sum_{S'_i \neq S_i}W_i(S_i \rightarrow S'_i)]
P''(S_1, \ldots , S_N; t)
\nonumber \\
+\sum_i [\sum_{S'_i \neq S_i}W_i(S'_i \rightarrow S_i)
P''(S_1, \ldots , S'_i, \ldots ,S_N; t)]\ .
\end{eqnarray}
Each spin can change from the value $S_i$ to the value $S'_i$ with 
probability per unit time 
\begin{equation}
W_i(S_i \rightarrow S'_i)=\frac{1}{\tau}
\frac{\exp(-\beta\Delta E''(S_i \rightarrow S'_i))}
{\sum_{S'_i}\exp(-\beta\Delta E''(S_i \rightarrow S'_i))}
\end{equation}
where the $\sum_{S'_i}$ is the sum over the 3 possible values of $S'_i$,
$\pm 1, 0$, and
\begin{equation}
\Delta E''(S_i \rightarrow S'_i)= 
-(S'_i-S_i)(J \sum_{<j>} \sigma_j + H) -({S'}_i^2-S_i^2)D
\end{equation}
gives the change in the energy of the system when the $S_i$ spin changes.
The probabilities satisfy the detailed balance condition
\begin{equation}
\frac{W_i(S_i \rightarrow S'_i)}{W_i(S'_i \rightarrow S_i)}
= \frac{P''_{eq}(S_1, \ldots , S'_i, \ldots ,S_N)}
{P''_{eq}(S_1, \ldots ,S_i, \ldots , S_N)}
\end{equation}
and, substituting the possible values of $S_i$, we get,
\begin{eqnarray}
W_i(1 \rightarrow \ 0)= W_i(-1 \rightarrow 0)=
\frac{\exp(-\beta D)}{2\cosh a + \exp(-\beta D)}
\nonumber \\
W_i(1 \rightarrow -1)= W_i(0 \rightarrow -1)=
\frac{\exp(-a)}{2\cosh a + \exp(-\beta D)}
\label{pput}\\
W_i(0 \rightarrow \ 1)= W_i(-1 \rightarrow 1)=
\frac{\exp(a)}{2\cosh a + \exp(-\beta D)}
\nonumber 
\end{eqnarray}
where $a= \beta (J \sum_{<j>} \sigma_j + H)$.
Notice that, since $W_i(S_i \rightarrow S'_i)$ does not depend on
$S_i$ we can write, $W_i(S_i \rightarrow S'_i)=W_i(S'_i)$, and the master 
equation becomes
\begin{eqnarray}
\frac{d}{dt}P''(S_1, \ldots , S_N; t)=
-\sum_i [\sum_{S'_i \neq S_i}W_i(S'_i)]
P''(S_1, \ldots ,S_N; t)
\nonumber \\
+\sum_i W_i(S_i)[\sum_{S'_i \neq S_i}
P''(S_1, \ldots , S'_i, \ldots , S_N; t)]
\label{master2}
\end{eqnarray}

Since the sum of probabilities is normalized to one,
by multiplying both sides of (\ref{master2}) by $S_k$ and taking
the average, we obtain
\begin{equation}
\tau \frac{d}{dt} <S_k>=-<S_k>+
\frac{2 \sinh \beta (J \sum_{<j>} \sigma_j + H)}
{2 \cosh \beta (J \sum_{<j>} \sigma_j + H) + \exp(-\beta D)}
\end{equation}
or, in terms of a mean-field approach
\begin{equation}
\tau \frac{d}{dt} <S>=-<S>+
\frac{2 \sinh \beta [J Z <\sigma> + H_0\cos(\omega t)]}
{2 \cosh \beta [J Z <\sigma> + H_0\cos(\omega t)] + \exp(-\beta D)}\ .
\label{m2}
\end{equation}

\bigskip
\leftline {\bf III. Results}

The system evolves according to the set of coupled differential equations
 given by (\ref{m1}) and (\ref{m2}), that can be written in the following form,
\begin{eqnarray}
\Omega \frac{d}{d\xi} m_1 = - m_1 + 
\frac{\sinh\frac{1}{T}(m_2+h\cos\xi)}
{\cosh\frac{1}{T}(m_2+h\cos\xi)+\frac{1}{2}\exp (-\beta D)}
\label{sys}\\
\Omega \frac{d}{d\xi} m_2 = - m_2 + 
\frac{1}{2}\tanh\frac{1}{2T}(m_1+h\cos\xi)
\nonumber 
\end{eqnarray}
where $m_1=<S>$ and $m_2=<\sigma>$, $\Omega=\tau\omega$, $\xi=\omega t$,
$T=(\beta J Z)^{-1}$ and $h=H_0/J Z$. We fix $J=-1$.
We are going to study the stationary solutions of this system and its
dependence with the parameters.

In general, the solutions of the system depend on the initial 
conditions, $m_1(\xi =0)$ and $m_2(\xi =0)$.
For a given set of parameters and initial conditions the system
pass through a transient regime until the solution becomes stationary. 
In Fig. 1, we show the solutions of the total magnetization,
$M=m_1+m_2$, for different initial conditions.
We found that the following combinations of initial conditions,
\begin{eqnarray}
m_1(\xi =0)=1,\ \ m_2(\xi =0)=0.5
\label{cia}\\
m_1(\xi =0)=0,\ \ m_2(\xi =0)=0\ \ 
\label{cib}
\end{eqnarray}
give all the possible different stationary solutions.
As expected, the sublattice magnetizations, $m_1$, $m_2$ and the total
magnetization $M$ are periodic functions of $\xi$ with period $2\pi$. 

By choosing as the dinamical parameter the average total magnetization 
in a period,
\begin{equation}
Q=\frac{1}{2\pi}\int_{0}^{2\pi}{M(\xi)d\xi}
\end{equation}
we can identify two types of solutions: a symmetric one where $M(\xi)$ 
follows the field, oscillating around zero
giving $Q = 0 $, and an antisymmetric one where $M(\xi)$
does not follows the field and oscillates around a finite value
different from zero, such that $Q \neq 0 $. Examples of both
types of solutions are shown in Fig. 2.
The first type of
solution is called paramagnetic (P) and the second type ferromagnetic (F).
The dynamical phase transition is located in the boundary between 
both solutions, and depending on the region of the phase diagram,
can be of first or second order. For the first order phase transition
the order parameter $Q$ is discontinuous, and jumps abruptly from
zero to a nonzero value.
For the second order phase transition, the order parameter decreases
continuously to zero, but its first derivative is discontinuous.
If we fix the values of $D$ and $\Omega$ for each value of $h$,
we get that the dynamical phase transition occurs at
temperature $T_c$.  In the thermodynamical plane, ($T$,$h$), we can define
a critical line $T_c=T_c(h)$, the point on the critical line at which
the transition changes from first order to second order
is denominated dynamical tricritical point.

As it is shown in Fig. 3, for $D$ large and positive,
we recover the phase
diagram of the standard Ising model~\cite{Tome}. At high temperatures the
solutions are paramagnetic and at low temperatures there are ferromagnetic.
The boundary between both regions, (F)$\rightarrow$(P), is given by the
critical line, and indicates a continuous phase transition. At low
temperatures, there is a range of values of $h$ for which the ferromagnetic
and paramagnetic solutions coexist and two separated critical lines
appear, one that indicates the discontinuous transition between the (F)
and the (P+F) regions, and the other that indicates the discontinuous
transition between the (P+F) and the (P) regions. The point where both
lines merge signals the change from a first order to a second order phase
transition and is a dynamical tricritical point.

When $D$ becomes negative, we found that in the range
$-2.03760\leq D<-0.11825$ (for $\frac{\Omega}{2\pi}=1$) and low
temperatures, a new first order critical line appears that separates
a new region where both solutions coexist (F+P) from the (F) region.
A phase diagram corresponding to a value of $D$ in this interval is
shown in Fig. 4.
This new critical line is due to the fact that, in this region, the
sublattice magnetization $m_1$ is very small, almost zero, such that
both sublattices are practically uncoupled, leaving $m_2$ free to
oscillate with the external field giving a paramagnetic type response,
$Q=0$, as seen in Fig. 5.
When $D \leq -2.03760$ the phase diagram changes dramatically. The
paramagnetic region is extended over the low $h$ regions, new critical
lines appear and a second dynamical tricritical point emerges. Phase
diagrams in this interval can be observed in Fig. 6.
The two first order transitions that occur at low temperature and the two
second order transitions at higher temperature are shown in Fig. 7.
In all cases we found that, as the crystalline field $D$ becomes larger
but negative, the region of the phase space where the system behaves
ferromagnetically becomes
narrower and shifts toward higher values of $h$. This effect, shown
in the inset of Fig. 6 is somewhat expected since,
for these values of $D$,
higher values of $h$ are needed to keep $m_1(\xi )$ different from zero
such that the order parameter, $Q$, can also be different from
zero. If $D$ is large enough and negative, the system behaves in almost
all the plane ($T$,$h$) as a paramagnet.

An interesting question arises at this point: is the existence of a second
tricritical point due to the fact that our model has a higher spin value,
$S=\pm 1,0$, than the Ising model, or to
the fact that the $S$ and $\sigma$ spin operators are alternated on the
lattice? To address this point, we study the Blume-Capel model (BC)
where a spin $S=\pm 1,0$ is located at each site of a square lattice.
The Hamiltonian of the BC model is
\begin{equation}
{\cal H} = -J \sum_{<nn>} S_i S_j - D \sum_i S_i^2
- H \sum_i S_i
\end{equation}
where $J$, $D$ and $H$ are in energy units. We choose $J=-1$ in order
to compare with our model.
Applying the Glauber-type stochastic dynamics and the mean-field
approach des\-cribed in II.b., we get that the time evolution of the BC
model is given by,
\begin{equation}
\Omega \frac{d}{d\xi} m = - m + 
\frac{\sinh\frac{1}{T}(m+h\cos\xi)}
{\cosh\frac{1}{T}(m+h\cos\xi)+\frac{1}{2}\exp (-\beta D)}
\label{blum}
\end{equation}
where $m=<S_i>$ is the total magnetization.

The numerical solution of (\ref{blum}) shows that, when
$D\ge 0$ the BC model has an Ising-type behavior: it has only one
dynamical tricritical point~\cite{Lee}. When $D$ is very small and
negative it behaves as our mixed Ising model: it
exhibits a new first order critical line for low temperatures, as
can be seen in Fig. 8.
However, as $D$ becomes larger but negative, the behavior of the BC model
departs completely from the mixed model: the second order critical line
disappears and, not only a new tricritical point does not appear, but 
the old one disappears, as can be seen in Fig. 9.
This study indicates that the existence of a second tricritical point
is not simply due to the fact that our model has a higher spin operator
but, to the mixing of the two types of spins.

All the results shown have been obtained with $\Omega/2 \pi=1.0$,
however we repeated our studies for different values of $\Omega/2 \pi$ and 
obtained qualitatively the same results. We found that the 
value of $D=-2.03760$, at which the system starts to exhibit two
dynamical tricritical points, is independent of the frequency, 
but the value of $D$ at which the model departs from the standard
Ising model by the appearance of a second critical line depends
on the frecuency as shown in Fig.10.
We found that, independently of the value of $D$, when $\Omega \rightarrow 0$
the region where both solutions coexist (P+F) vanishes and the tricritical
temperature approaches the static critical value.

\bigskip
\leftline {\bf IV. Conclusions}

We have analyzed within a mean-field approach the kinetics of a classical
mixed Ising ferrimagnetic model in the presence of a time-dependent
oscillating external field. We use a Glauber-type stochastic dynamics to
describe the time evolution of the system.
We found that the behavior of the system strongly depends on the values of
the crystal field parameter, $D$. For large and positive values of $D$ the
system behaves as the standard Ising model~\cite{Tome}, it has a continuous
phase transition at high $T$ and low $h$. As $T$ decreases and $h$ increases
the transition becomes discontinuous and the system presents a dynamical
tricritical point, see Fig. 3. However, when $D$ is large
enough and negative the
phase diagram changes completely. First, a new critical line appears at low
$T$ that marks a discontinuous phase transition between a new coexistence
region (P+F) and the (F) region. As $D$ becomes larger (and negative) the
ferromagnetic region shrinks, the paramagnetic region also covers the low
$h$ region, new critical lines appear and a second dynamical tricritical
point emerges, see Fig. 6. The minimum value of $|D|$ at which
the new tricritical point appears seems to be independent of $\Omega $
(for $\Omega \neq 0$).

A similar study on the Blume-Capel model ($S=\pm 1,0$) reveals that the
behavior of our mixed ferrimagnetic model, in particular the appearance of
two tricritical points, is intimately related to the mixing of the two
types of spin operators, and not only due to the fact of having a higher
spin operator.

This mean-field study suggests that the mixed Ising ferrimagnetic model has
an interesting dynamical behavior, quite different from the standard Ising
model, and it would be worth to further explore it with more accurate
techniques such as Monte Carlo simulations or renormalization group
calculations.

\bigskip
\noindent {\bf Acknowledgments:}

We are grateful to M. A. Novotny and P. A. Rikvold for useful discussions.

\newpage
\begin{center} {\bf Figure Captions} \end{center}
\bigskip

\noindent {\bf Fig. 1.} Solutions for different initial conditions with
$T=0.05$, $h=0.5$, $\Omega/2\pi=1.0$ and $D=20$. As we see, after the
transient regime, some different initial conditions give the same solutions.
\bigskip

\noindent {\bf Fig. 2.} Types of solutions: antisymmetric where $M$ oscillates
around a finite value different from zero, and symmetric where $M$ oscillates
around zero. For $T=0.05$, $h=0.5$, $\Omega/2\pi=1.0$ and $D=20$.
\bigskip

\noindent {\bf Fig. 3.} Phase diagram in the ($T$,$h$) plane for
$\Omega /2\pi =1.0$ and $D=20.0$. The paramagnetic (P) and the ferromagnetic
(F) solutions overlap in the region indicated by P+F. The $\bigcirc$ symbol
indicates the dynamical tricritical point.
\bigskip

\noindent {\bf Fig. 4.} Phase diagram in the plane ($T$,$h$) for
$\Omega /2\pi =1.0$ and $D=-1.5$. The $\bigcirc$ symbol indicates the dynamical
tricritical point. Notice the appearance of a new critical line at low $T$.
\bigskip

\noindent {\bf Fig. 5.} Behaviour of $m_{1}(\xi )$ and $m_{2}(\xi )$ for
$T=0.05$, $h=0.2$, $\Omega /2\pi =1.0$ and $D=-1.5$.
\bigskip

\noindent {\bf Fig. 6.} Phase diagram in the plane ($T$,$h$) for
$\Omega /2\pi =1.0$ and $D=-2.5$. In the inset we compare the phase diagram
for two values of $D$: $D_1=-2.5$ and $D_2=-2.7$.
Again the $\bigcirc$ symbols indicates the dynamical tricritical points.
\bigskip

\noindent {\bf Fig. 7.} Order parameter $Q$ as function of the external
magnetic field $h$ for $\Omega /2\pi =1.0$ and $D=-2.5$;
(a) for $T=0.0025$ and (b) for $T=0.0305$.
\bigskip

\noindent {\bf Fig. 8.} Phase diagram of the Blume-Capel model in the plane
($T$,$h$) for $\Omega /2\pi =1.0$ and $D=-1.0$.
Again the $\bigcirc$ symbol indicates the dynamical tricritical point.
\bigskip

\noindent {\bf Fig. 9.} Phase diagram of the Blume-Capel model in the plane
($T$,$h$) for $\Omega /2\pi =1.0$ and $D=-2.5$. In the inset we compare the
phase diagram for two values of $D$: $-2.5$ and $-3.0$.
\bigskip

\noindent {\bf Fig. 10.} Dependence between the minimum value of $D$ at which
the system departs from the standard Ising model by the appearance of other
critical line, and the frecuency $\Omega /2\pi$.

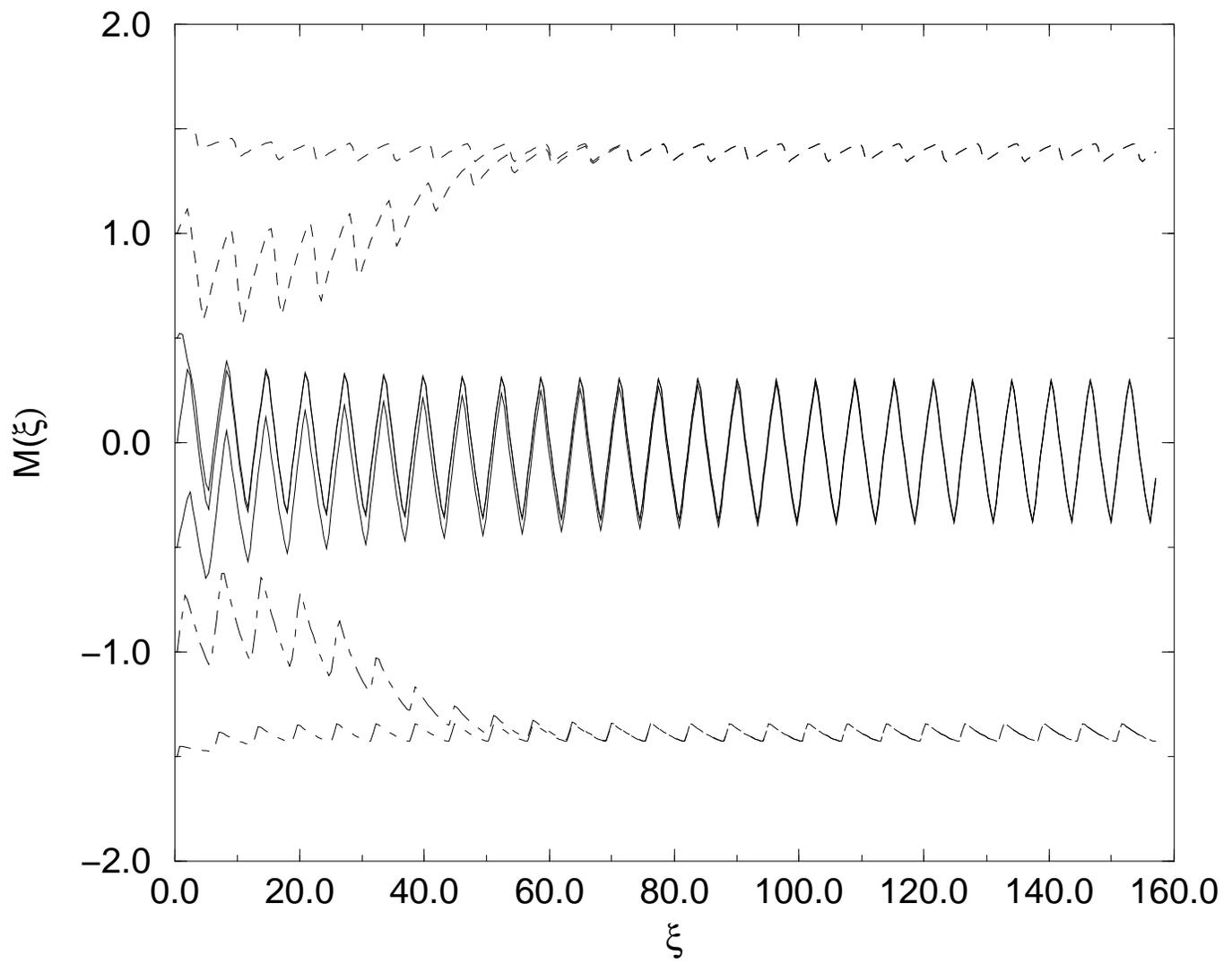
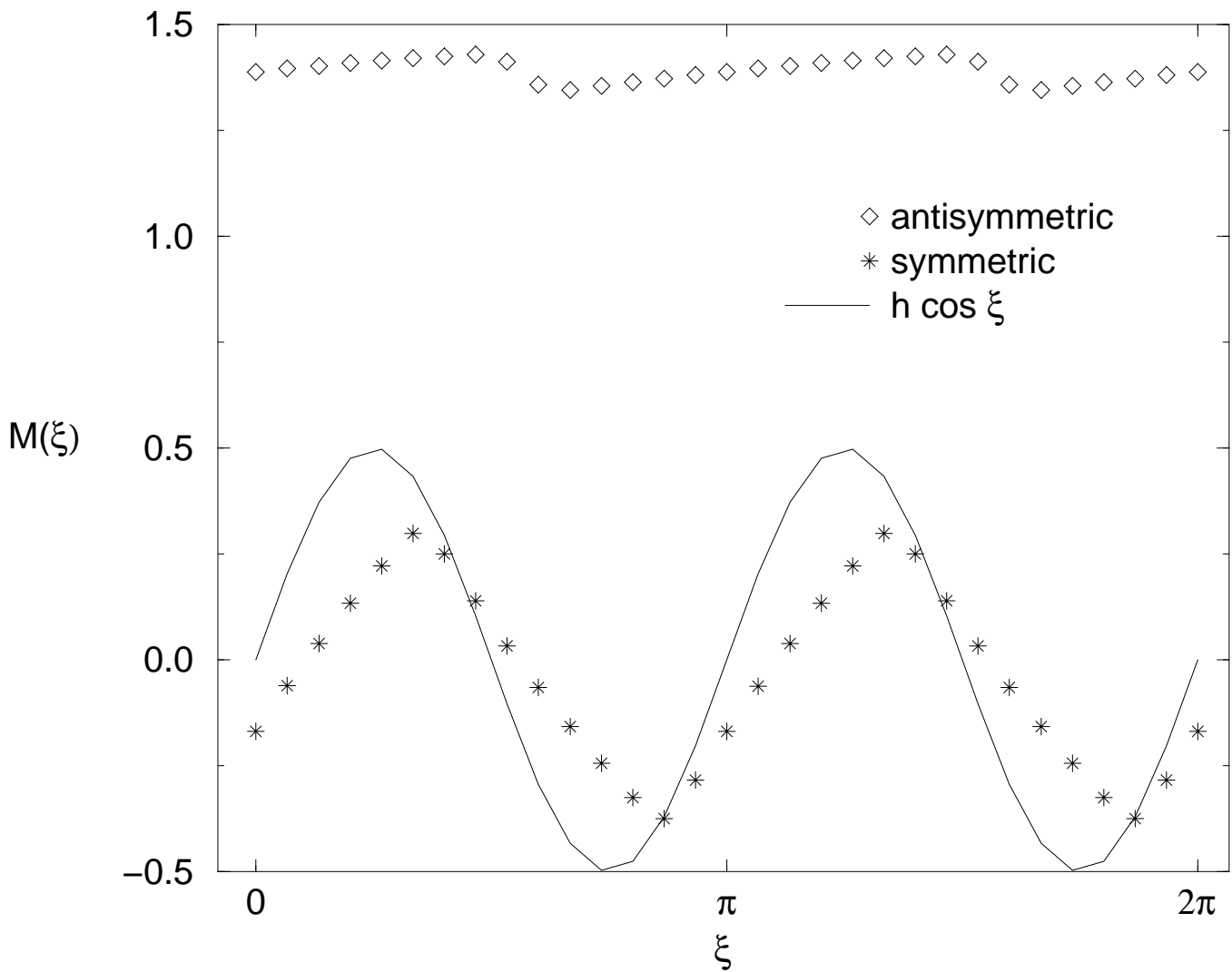
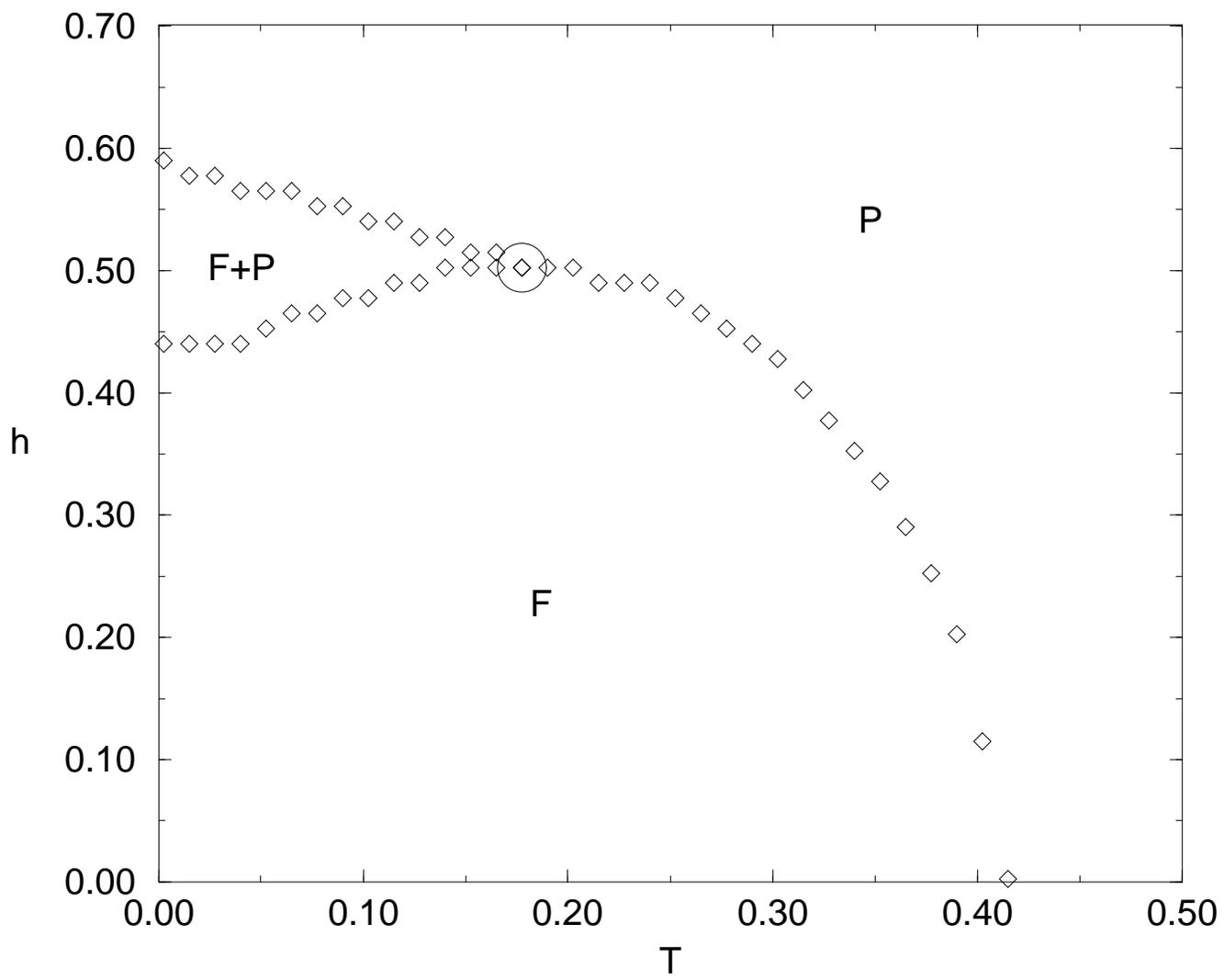
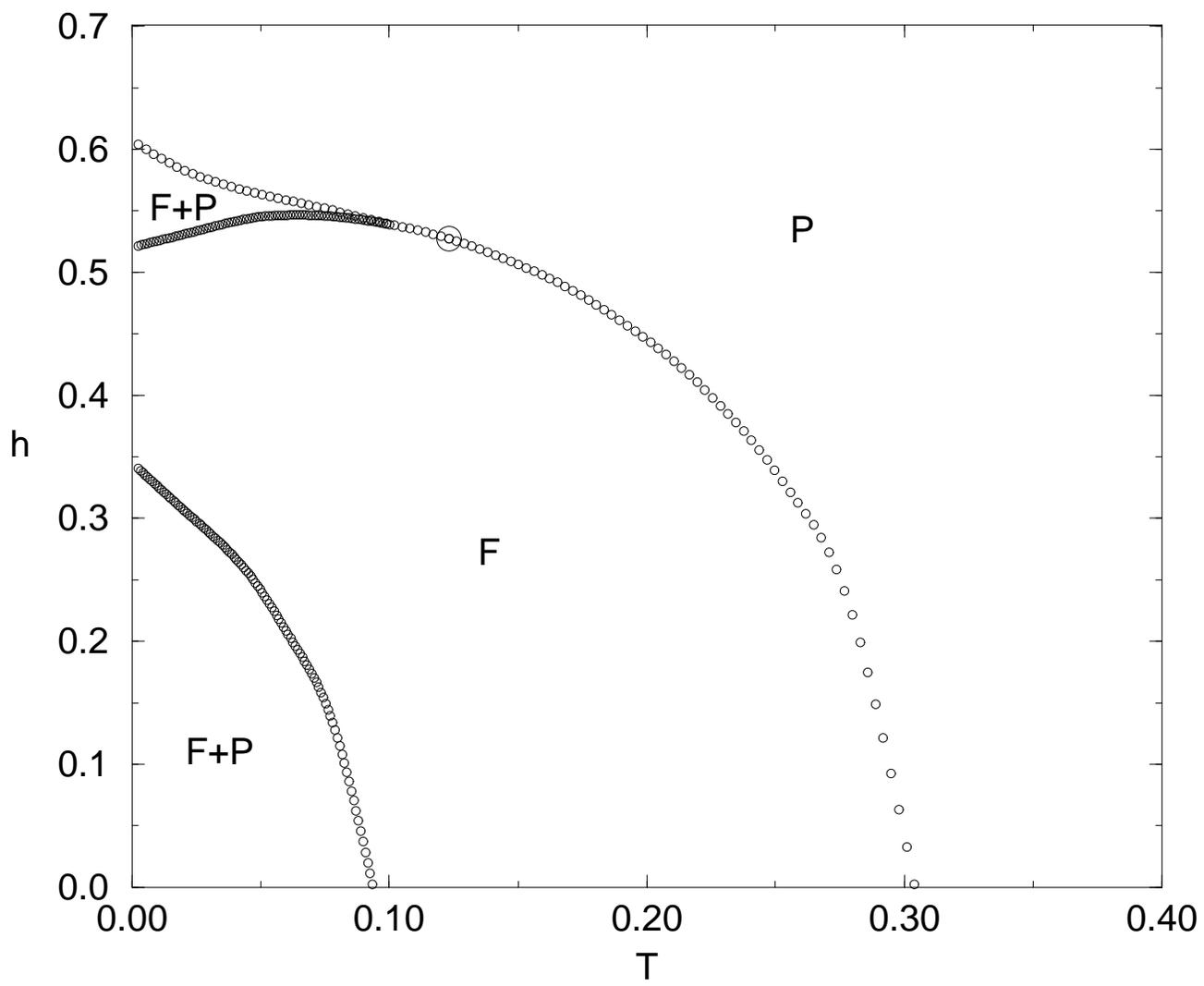
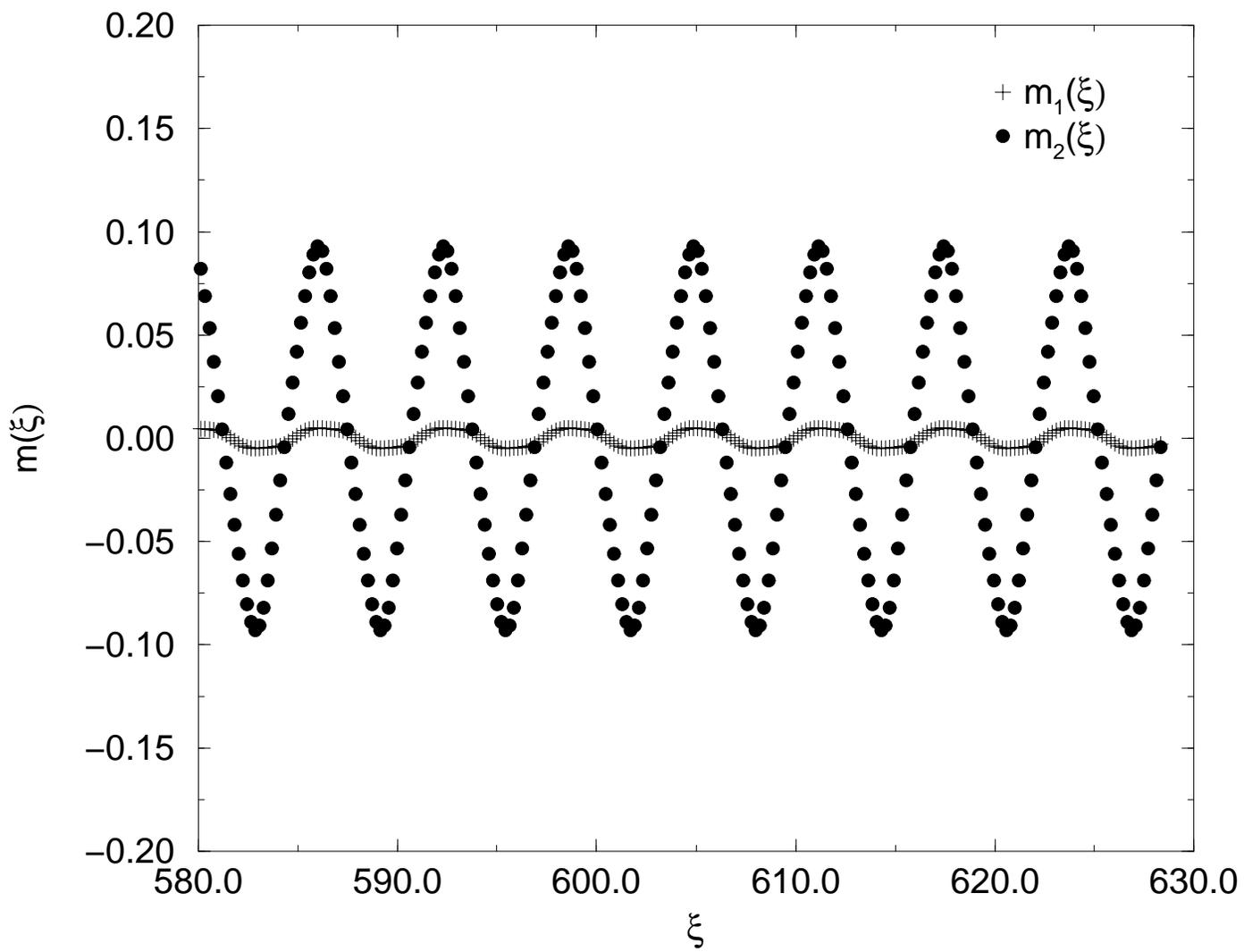
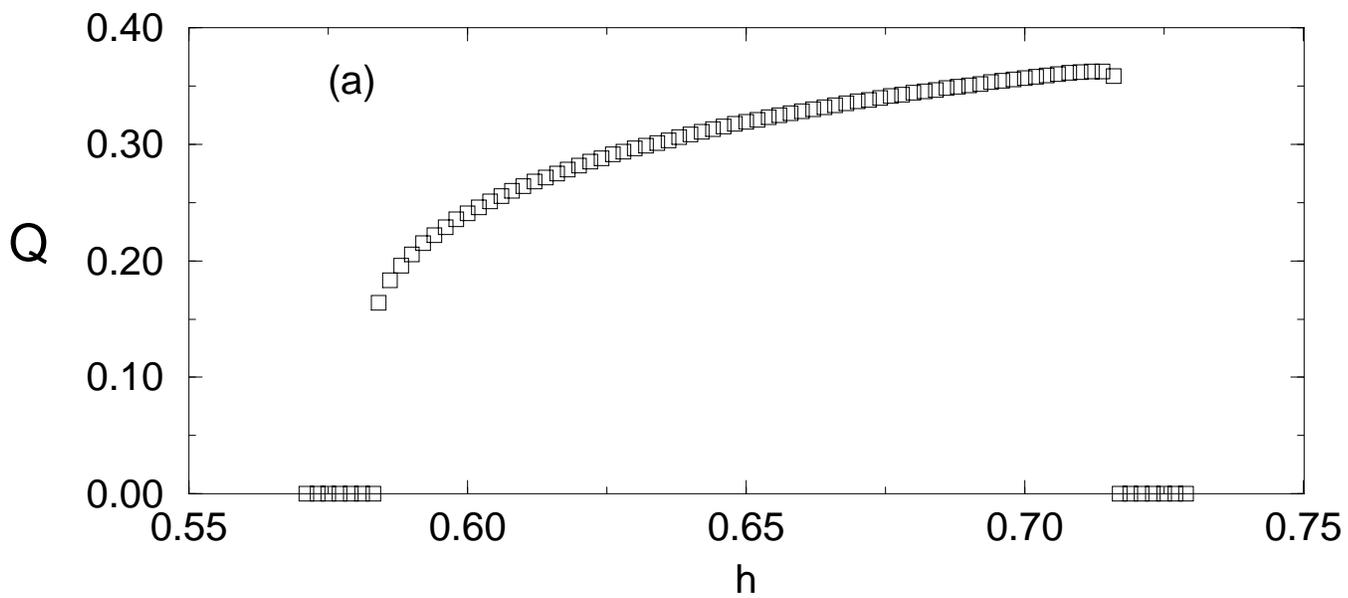
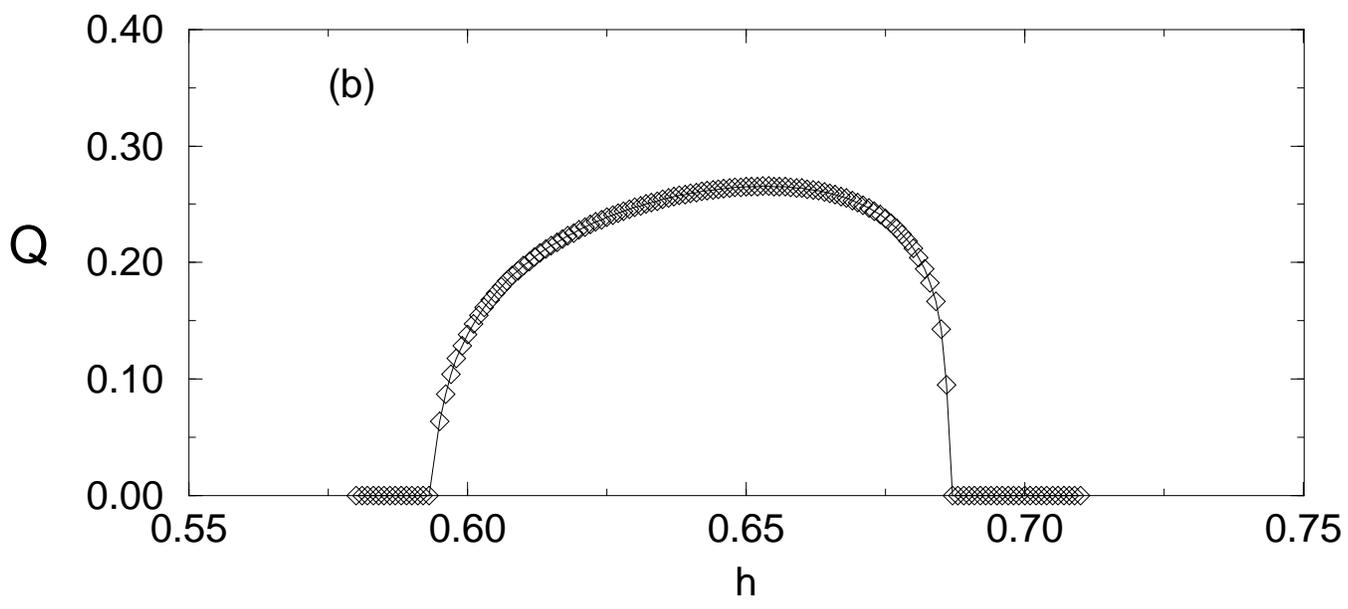
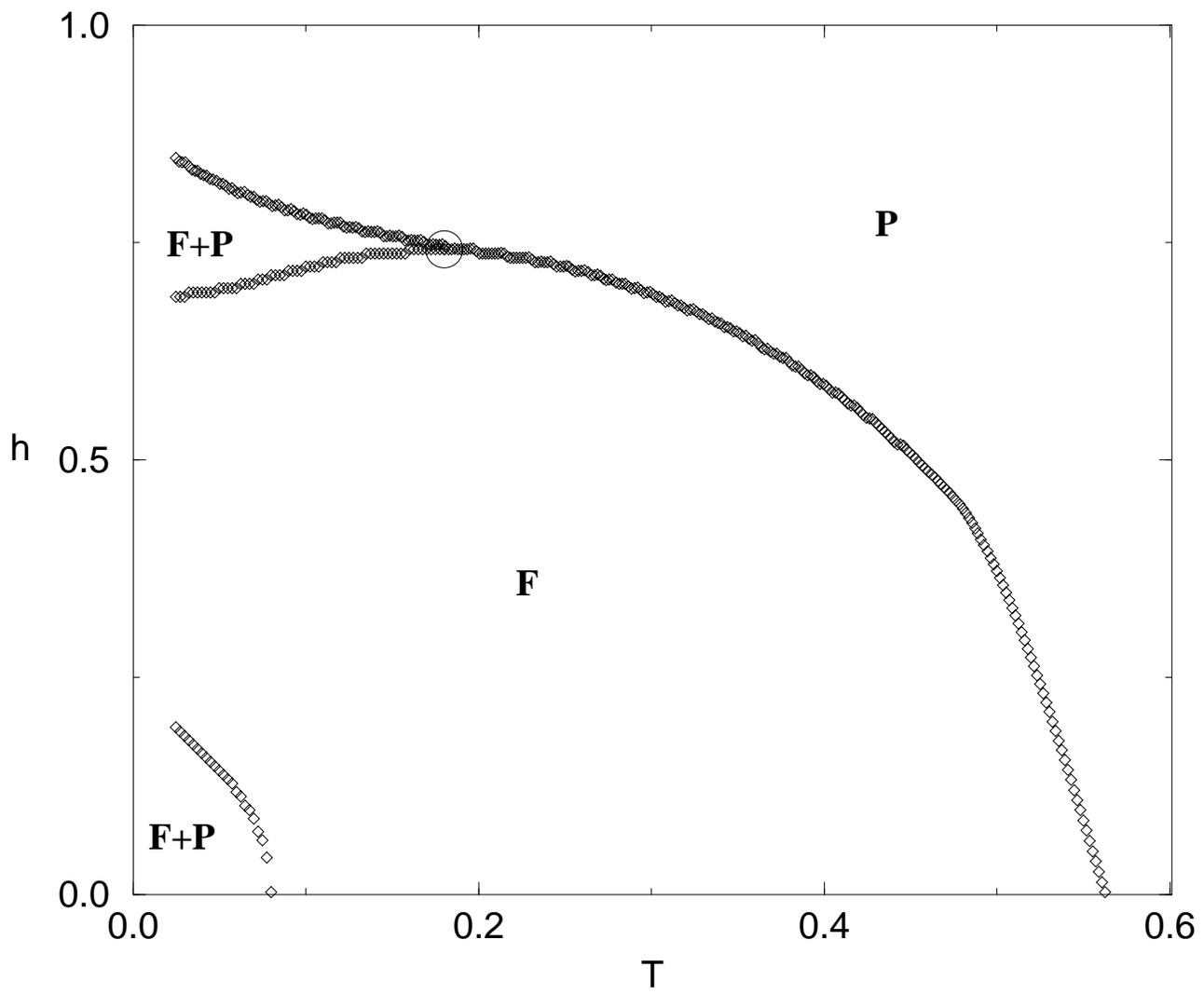
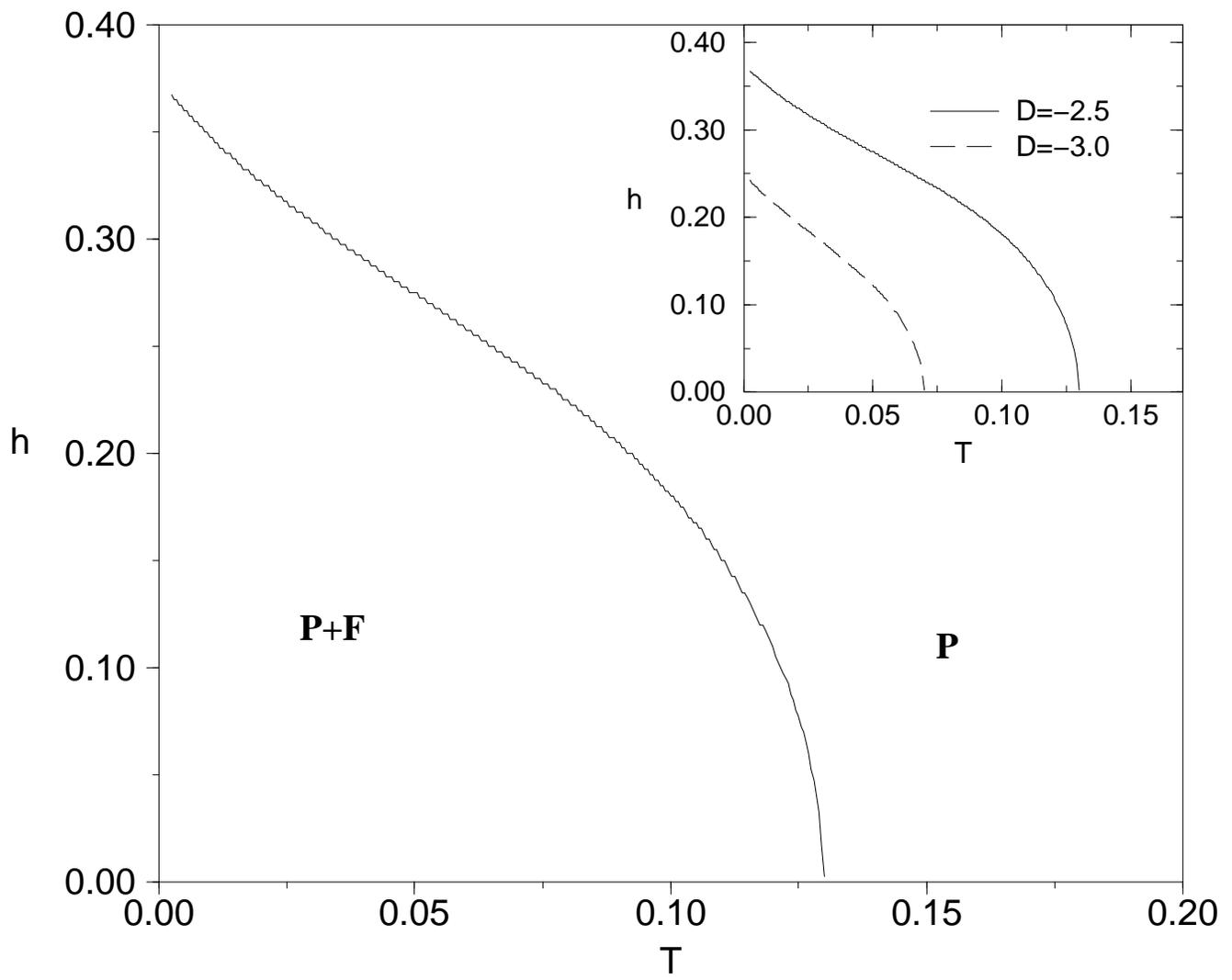
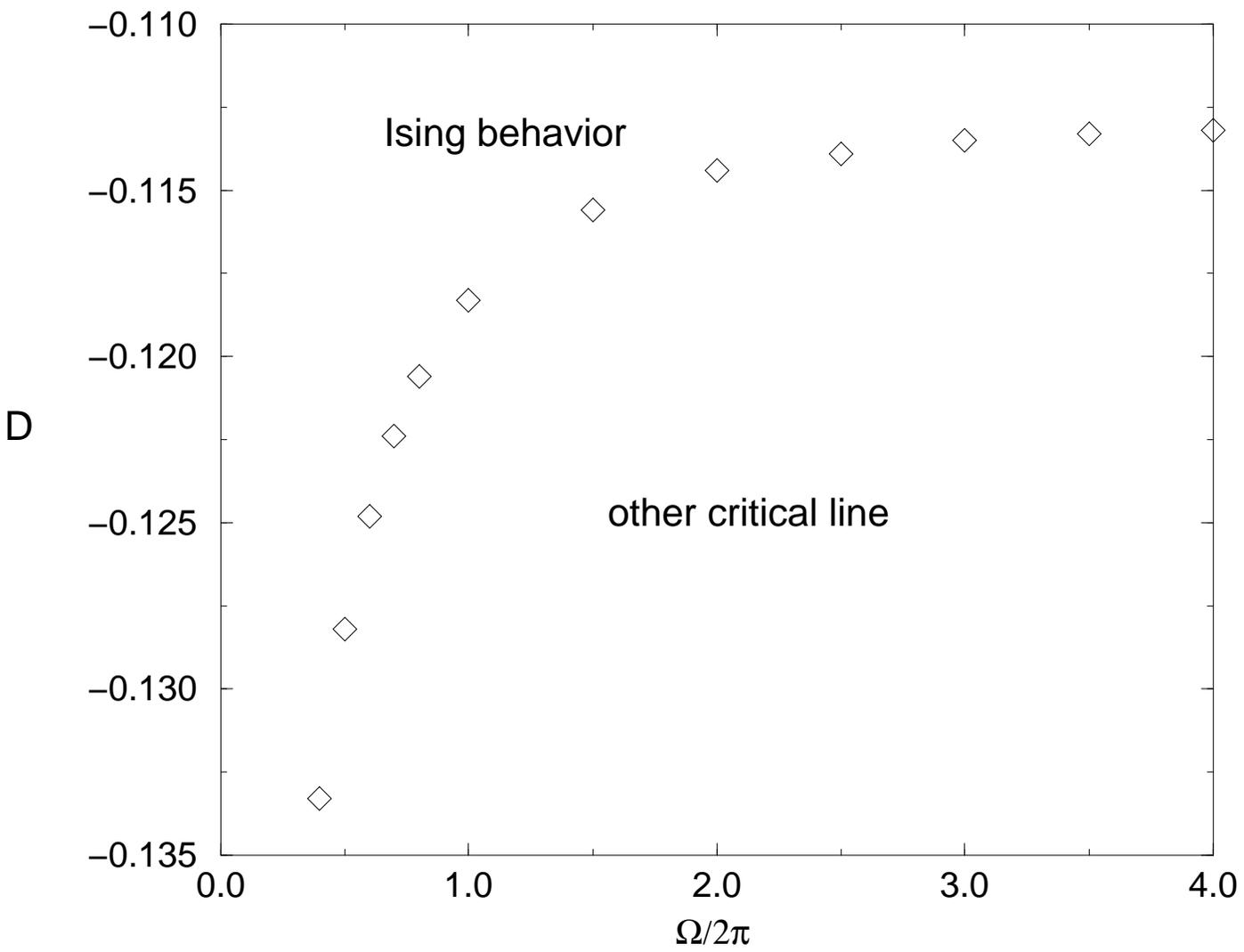
\end{document}